\documentclass{article}
\usepackage{amsfonts}
\usepackage{amsmath}
\usepackage{mathtext}
\usepackage[english]{babel}
\usepackage[dvips]{graphicx}

\def\be{\begin{equation}}\def\ee{\end{equation}}\def\l{\label}
\def\0{\setcounter{equation}{0}}
\def\ba{\begin{eqnarray}}\def\ea{\end{eqnarray}}

\def\s{\sigma}\def\f{\frac}\def\pa{\partial}

\def\foot{\footnote}\def\vep{\varepsilon}

\def\ba{\mathbf a}

\begin{document}

\begin{center}
{\Large\bf A note about the t`Hooft`s ansatz for $SU(N)$ real time guage
theories.}
\vglue 0.1cm
J.Manjavidze\foot{Inst. Phys. (Tbilisi, Georgia) $\&$ JINR 
(Dubna, Russia), E-mail: joseph@nu.jinr.ru},
V.Voronyuk\foot{JINR (Dubna, Russia), E-mail: vadimv@nu.jinr.ru}
\vglue 0.1cm
JINR, Dubna
\end{center}

The t`Hooft's ansatz reduces the classical Yang--Mills theory to the
$\lambda\phi^4$ one. It is shown that in the frame of this ansatz
the real-time classical solutions for the arbitrary $SU(N)$ gauge
group is obtained by embedding $SU(2)\times SU(2)$ into
$SU(N)$. It is argued that this group structure is the only
possibility in the frame of the considered ansatz. New explicit
solutions for $SU(3)$ and $SU(5)$ gauge groups are shown.

\section{Introduction}

In order to simplify the problem of solving a Yang--Mills equation
for the vector field it was offered by t`Hooft {\it at. al.} the ansatz for the
Euclidean space \cite{Hooft}. It reduces the Yang--Mills equation
to the equation for a single scalar field $\phi$. The $SU(2)$
classical solutions discovered by means of this ansatz are well
known \cite{Actor} and were used to generate $SU(N)$ solutions
by simply embedding $SU(2)$ into $SU(N)$ \cite{Bernard}.

One of them allows the coordinate transformation to the Minkowski space
so that it becomes nonsingular, real and possesses a finite action
and energy \cite{deAlfaro:1976,Actor}.

The $SU(2)$ gauge group was assumed for both the Euclidean and
Minkowski space, see also \cite{Bernreuther}, while the
experimental analysis shows that QCD is the $SU(3)$ gauge theory
\cite{Bar}. So, the knowledge of the real-time classical solution for QCD
is important since it allows to analyze the non-perturbative
corrections \cite{Soso} to the observables.

In this article we will try to find a $SU(N)$
solution by means of the t`Hooft`s ansatz. The only condition we
assume for the ansatz is the following: it must reduce the Yang--Mills
equation to the real scalar $\lambda\phi^4$ theory. We will solve
this condition and will show that the only solution of the classical
Yang--Mills equation in the frame of the t'Hooft's ansatz is embedding
$SU(2)\times SU(2)$ into $SU(N)$.

\section{Definition of ansatz}

Let us start from the Yang--Mills equation in the matrix form \be
\pa^\mu F_{\mu \nu}+i g  [A^{\mu}, F_{\mu \nu}]=0, \l{a}\ee where
$$A_\mu=t_a A_{a\mu},$$
$$F_{\mu \nu}=
\pa_\mu A_{\nu}-\pa_\nu A_{\mu}+i g [A_{\mu}, A_{\nu}],$$
$t_a$ are generators of the gauge group.

Let us consider the t`Hooft`s ansatz without any assumptions about
gauge group
$$
A_{\mu}(x)=\f{1}{g}\eta_{\mu \nu}\,\pa^\nu \ln \phi (x),
$$
where $\eta_{\mu \nu}$ are some matrixes. We will consider that
$A_{\mu}(x)$ satisfies the Lorentz gauge condition: $\pa^\mu
A_{\mu}=0$ and so $\eta_{\mu \nu}$ are antisymmetric over $\mu$
and $\nu$ matrixes. It is assumed that $\eta_{\mu \nu}$ are
constant in this gauge.

It is necessary to take the equality \be\label{eta:ysl}
-i[\eta_{\mu \s},\eta_{\nu \rho}]= \eta_{\mu \nu}g_{\rho \s}-
\eta_{\mu \rho}g_{\s \nu}+ \eta_{\s \rho}g_{\mu \nu}- \eta_{\s
\nu}g_{\mu \rho} \ee in order to reduce the Yang--Mills equation to
the equation for the single scalar field. As the result of
substitution of ansatz with the property (\ref{eta:ysl}) into 
Yang--Mills equation (\ref{a}), we have
\be\label{phi4} \Box\phi+\lambda\phi^3=0, \ee where $\lambda$ is
an arbitrary integration constant. Emphasize that the
Eq.(\ref{phi4}) is the result of (\ref{eta:ysl}), this reduction is
valid for any gauge group.

Therefore, the problem (\ref{a}) was divided into two parts:
the searching of $\eta_{\mu \nu}$ from the algebraic equality
(\ref{eta:ysl}) and the solving of the equation (\ref{phi4}) for
$\phi(x)$.

Particular solutions of the equation (\ref{phi4}) are known, see
\cite{Actor,deAlfaro:1976,Castell,Schechter}, and we will not
consider this question.

The matrices $\eta_{\mu\nu}$ can be written in a convenient form
\be\label{eta:def} \eta_{\mu\nu}=-\vep_{0\mu\nu\kappa}X_{\kappa}+
i g_{0\mu} Y_{\nu} - i g_{0\nu} Y_{\mu}, \qquad \kappa=1,2,3, \ee
since they are antisymmetric, where $\vep_{0123}=1$; the unknown
$X_i$ and $Y_i$ are matrixes in the group space, $X_0=0$, $Y_0=0$,
$X_i=-X^i$, $Y_i=-Y^i$.

Let us insert  (\ref{eta:def}) into (\ref{eta:ysl}). Then we
obtain algebraic equations for $X_i$ and $Y_i$. Because of
antisymmetry of $\eta_{\mu\nu}$, it is convenient to examine only
three cases

1. $\mu=0, \s=i, \nu=0, \rho=j$, where $i,j=1,2,3$.
Then we have
\be\label{ysl:1}
[Y_i,Y_j]=i \vep_{i j k} X_k;
\ee

2. $\mu=0, \s=i, \nu=j, \rho=k$, where $i,j,k=1,2,3$. It is easy
to obtain
$$
\vep_{j k s} [Y_i,X_s]=i Y_{j} g_{ik}-i Y_{k} g_{ij}.
$$
So, we have:
\begin{equation}\label{ysl:2}
[Y_i,X_j]=i \vep_{i j k} Y_k;
\end{equation} after changing the indexes

3. $\mu=i, \s=j, \nu=k, \rho=s$, where $i,j,k,s=1,2,3$.
This case gives
$$
-i[(-\vep_{ijp} X_p),(-\vep_{ksl} X_l)]
=
(-\vep_{ikp} X_p)g_{sj}-(-\vep_{isp} X_p)g_{jk}+
(-\vep_{jsp} X_p)g_{ik}-(-\vep_{jkp} X_p)g_{is}
$$
After symplification and changing indexes we have
\be\label{ysl:3}
[X_i,X_j]=i \vep_{i j k} X_k.
\ee
The other cases can be easily reduced to this three ones.

It follows from (\ref{ysl:1},\ref{ysl:2},\ref{ysl:3}) that \be
\left[\mathcal J_i,\mathcal J_j\right] = i \vep_{i j k} \mathcal
J_k \qquad \left[\mathcal K_i, \mathcal K_j\right] = i \vep_{i j
k} \mathcal K_k, \l{b}\ee
$$
\left[\mathcal J_i, \mathcal K_j\right] =0,
$$
where
$$
\mathcal J_i=\f{X_i+Y_i}2, \qquad \mathcal K_i=\f{X_i-Y_i}2.
$$ It follows from (\ref{b}) that $N\times N$ matrixes
$\mathcal J_i$ and $\mathcal K_i$ are elements of the $SU(2)\times
SU(2)$ group. Then the ansatz can be written as follows
\be\label{eta:sol} \eta_{\mu\nu}= \left(-\vep_{0\mu\nu\kappa}
\mathcal J_{\kappa}+ i g_{0\mu} \mathcal J_{\nu} - i g_{0\nu}
\mathcal J_{\mu}\right)+ \left(-\vep_{0\mu\nu\kappa}\mathcal
K_{\kappa}- i g_{0\mu} \mathcal K_{\nu} + i g_{0\nu} \mathcal
K_{\mu}\right), \qquad \kappa=1,2,3. \ee

This is the general solution of (\ref{eta:ysl}) and, therefore,
it is unique. There always exists a nonzero t'Hooft's ansatz for any $N
\ge 2$ since the representation of the $SU(2)\times SU(2)$ group by
$N\times N$ matrixes always exists. The meaning of such
representation is embedding $SU(2)\times SU(2)$ into $SU(N)$.

This ansatz gives complex potentials $A_\mu$ for real $\phi$,
however one can check that it leads to a real Lagrangian density.
Therefore one can expect that there exists some complex gauge
transformation which makes it real as it was done for $SU(2)$
\cite{deAlfaro:1976}.

Let us consider the solutions for $SU(2)$, $SU(3)$ and $SU(5)$
groups.

\subsection{SU(2)}
For the $SU(2)$ gauge group the only solution is (either $\mathcal
J_i$ or $\mathcal K_i$ is equal to zero)
$$
X_i=\pm Y_i=\f{\s_i.}{2}
$$
Then we obtain well-known solution \cite{Hooft,Actor} which can be
written in a component form:
$$
\eta_{a\mu\nu}=-\vep_{0a\mu\nu}\mp i g_{0\mu}g_{a\nu}
\pm i g_{0\nu}g_{a\mu}.
$$

\subsection{SU(3)}
For the $SU(3)$ gauge group also either $\mathcal J_i$ or
$\mathcal K_i$ is equal to zero, so we have
$$
X_i=\pm Y_i.
$$
There exists both reducible and irreducible representation of the
$SU(2)$ group in terms of $3\times3$ matrixes.

\subsubsection{Reducible representation}
The $SU(3)$ group contains $3$ independent $SU(2)$ subgroups
which are not form direct product. So there exist $3$ independent
solutions:

\noindent
{\bf (I)}: $X_{1}^{(I)}=t_1, X_{2}^{(I)}=t_2, X_{3}^{(I)}=t_3$

In the component form we obtain
$$
\eta_{1\,\mu\nu}=
\left(
\begin{array}{cccc}
0 & \pm i & 0 & 0 \\
\mp i & 0 & 0 & 0 \\
0 & 0 & 0 & -1 \\
0 & 0 & 1 & 0
\end{array}
\right)_{\mu\nu},
\quad
\eta_{2\,\mu\nu}=
\left(
\begin{array}{cccc}
0 & 0 & \pm i & 0 \\
0 & 0 & 0 & 1 \\
\mp i & 0 & 0 & 0 \\
0 & -1 & 0 & 0
\end{array}
\right)_{\mu\nu},
\quad
\eta_{3\,\mu\nu}=
\left(
\begin{array}{cccc}
0 & 0 & 0 & \pm i \\
0 & 0 & -1 & 0 \\
0 & 1 & 0 & 0 \\
\mp i & 0 & 0 & 0
\end{array}
\right)_{\mu\nu},
$$
$$
\eta_{a\,\mu\nu}=0, a=4,...,8.
$$
\noindent
{\bf (II)}: $X_1^{(II)}=t_4, X_2^{(II)}=t_5,
X_3^{(II)}=\f12(\sqrt{3}\, t_8+t_3);$

\noindent
{\bf (III)}: $X_1^{(III)}=t_6, X_2^{(III)}=t_7,
X_3^{(III)}=\f12(\sqrt{3}\, t_8-t_3).$

The cases (II) and (III) are similar to the (I) with the
difference in gauge indexes.

\subsubsection{Irreducible representation}
There also exist irreducible representation of the $SU(2)$ group
by $3\times 3$ matrixes.
$$
X_1=\f{1}{\sqrt{2}}
\left(
\begin{array}{ccc}
0 & 1 & 0 \\
1 & 0 & 1 \\
0 & 1 & 0
\end{array}
\right),
\quad
X_2=\f{1}{\sqrt{2}}
\left(
\begin{array}{ccc}
0 & -i & 0 \\
i & 0 & -i \\
0 & i & 0
\end{array}
\right),
\quad
X_3=
\left(
\begin{array}{ccc}
1 & 0 & 0 \\
0 & 0 & 0 \\
0 & 0 & -1
\end{array}
\right).
$$
Then in the component form we obtain
$$
\eta_{1\,\mu\nu}= \sqrt{2}
\left(
\begin{array}{cccc}
0 & \pm i & 0 & 0 \\
\mp i & 0 & 0 & 0 \\
0 & 0 & 0 & -1 \\
0 & 0 & 1 & 0
\end{array}
\right)_{\mu\nu},
\quad
\eta_{2\,\mu\nu}= \sqrt{2}
\left(
\begin{array}{cccc}
0 & 0 & \pm i & 0 \\
0 & 0 & 0 & 1 \\
\mp i & 0 & 0 & 0 \\
0 & -1 & 0 & 0
\end{array}
\right)_{\mu\nu},
\quad
\eta_{3\,\mu\nu}=
\left(
\begin{array}{cccc}
0 & 0 & 0 & \pm i \\
0 & 0 & -1 & 0 \\
0 & 1 & 0 & 0 \\
\mp i & 0 & 0 & 0
\end{array}
\right)_{\mu\nu},
$$
$$
\eta_{4\,\mu\nu}=\eta_{5\,\mu\nu}=0,
\quad
\eta_{6\,\mu\nu}=\eta_{1\,\mu\nu},
\quad
\eta_{7\,\mu\nu}=\eta_{2\,\mu\nu},
\quad
\eta_{8\,\mu\nu}=\sqrt{3} \:\eta_{3\,\mu\nu}.
$$

\subsection{SU(5)}
Considering the $SU(5)$ group it is interesting to examine the
solution with both nonzero $SU(2)$ groups. If $\mathcal J_i$ or
$\mathcal K_i$ is equal to zero then the solution will be given by
reducible or irreducible representation of the group in a way like
$SU(3)$.

For the $\mathcal J_i$ one can take irreducible group presentation
for the $3\times3$ matrixes in the upper left corner and for the
$\mathcal K_i$ one can take $2\times2$ group presentation for the lower
right corner, and vice versa. It can be written in the obvious
form:
$$
\left(
\begin{array}{cc}
\begin{array}{c}
\mathcal J\\SU(2)\\ 3\times3
\end{array}&
\begin{array}{cc}
0&0\\
0&0\\
0&0
\end{array}
\\
\begin{array}{ccc}
0&0&0\\
0&0&0
\end{array} &
\begin{array}{c}
\mathcal K\\SU(2)\quad 2\times2
\end{array}
\end{array}
\right)
$$
Then the ansatz in component form is as follows:
$$
\eta_{1\,\mu\nu}= \sqrt{2}
\left(
\begin{array}{cccc}
0 & \pm i & 0 & 0 \\
\mp i & 0 & 0 & 0 \\
0 & 0 & 0 & -1 \\
0 & 0 & 1 & 0
\end{array}
\right)_{\mu\nu},
\quad
\eta_{2\,\mu\nu}= \sqrt{2}
\left(
\begin{array}{cccc}
0 & 0 & \pm i & 0 \\
0 & 0 & 0 & 1 \\
\mp i & 0 & 0 & 0 \\
0 & -1 & 0 & 0
\end{array}
\right)_{\mu\nu},
\quad
\eta_{3\,\mu\nu}=
\left(
\begin{array}{cccc}
0 & 0 & 0 & \pm i \\
0 & 0 & -1 & 0 \\
0 & 1 & 0 & 0 \\
\mp i & 0 & 0 & 0
\end{array}
\right)_{\mu\nu},
$$
$$
\eta_{4\,\mu\nu}=\eta_{5\,\mu\nu}=0,
\quad
\eta_{6\,\mu\nu}=\eta_{1\,\mu\nu},
\quad
\eta_{7\,\mu\nu}=\eta_{2\,\mu\nu},
\quad
\eta_{8\,\mu\nu}=\sqrt{3} \:\eta_{3\,\mu\nu},
\quad
\eta_{9...20\,\mu\nu}=0,
$$
$$
\eta_{21\,\mu\nu}=
\left(
\begin{array}{cccc}
0 & \mp i & 0 & 0 \\
\pm i & 0 & 0 & 0 \\
0 & 0 & 0 & -1 \\
0 & 0 & 1 & 0
\end{array}
\right)_{\mu\nu},
\quad
\eta_{22\,\mu\nu}=
\left(
\begin{array}{cccc}
0 & 0 & \mp i & 0 \\
0 & 0 & 0 & 1 \\
\pm i & 0 & 0 & 0 \\
0 & -1 & 0 & 0
\end{array}
\right)_{\mu\nu},
\quad
\eta_{23\,\mu\nu}=
\left(
\begin{array}{cccc}
0 & 0 & 0 & \mp i \\
0 & 0 & -1 & 0 \\
0 & 1 & 0 & 0 \\
\pm i & 0 & 0 & 0
\end{array}
\right)_{\mu\nu},
$$
$$
\eta_{24\,\mu\nu}=0.
$$
If one believes that the $SU(5)$ theory is unification of
electro--weak and strong interactions then indexes $a=1,...,8$
correspond to the strong and $a=21,...,23$ to the electro--weak
interactions. But one can see that this solution can not be used
for this purpose.

\section{Conclusions}

In the frame of the ansatz the $SU(N)$ classical solutions
always exist and each one is given by embedding $SU(2)\times
SU(2)$ into $SU(N)$.

Let us assume that $\phi$ is invariant under $O(4)\times O(2)$
coordinate transformations \cite{deAlfaro:1976,Schechter}. In the
frame of this prescription, we obtain the real solution of the
Yang-Mills equation
$$
A_0=
\pm \f{x_0 x_a}{g\: y^2} \mathcal J_a
\mp \f{x_0 x_a}{g\: y^2} \mathcal K_a,
$$
$$
A_i=
\f{1}{g\:y^2}\left[
-\vep_{a i n}x_n \pm \delta_{a i}\f12 (1+x^2)\pm x_a x_i
\right] \mathcal J_a +
\f{1}{g\:y^2}\left[
-\vep_{a i n}x_n \mp \delta_{a i}\f12 (1+x^2)\mp x_a x_i
\right] \mathcal K_a,
$$
where
$$
y^2=\f14 (1-x^2)^2+x_0^2,
\qquad
\vep_{1 2 3}=1, \quad n=1...3,
$$
and $\mathcal J_a$, $\mathcal K_a$ are corresponding representation
of $SU(2)\times SU(2)$ group by $N\times N$ matrixes.

\vglue 1em

{\bf Acknowledgement}

The authors thank A.Dorokhov for the useful discussion. One of the authors
(V.Voronyuk) is indebted to Prof. R.Kragler for his help and to
DAAD for the financial support.

\end{document}